# Scaling Behavior of the Quantum Phase Transition from a Quantum-Anomalous-Hall Insulator to an Axion Insulator

Xinyu Wu[1], Di Xiao[2], Chui-Zhen Chen[3], Jian Sun[1], Ling Zhang[2], Moses H. W. Chan[2], Nitin Samarth[2], X. C. Xie[1,4,5], Xi Lin[1,4,5], and Cui-Zu Chang[2]

[1]International Center for Quantum Materials, Peking University, Beijing 100871, China

[2]Department of Physics, The Pennsylvania State University, University Park, PA 16802, USA

[3]Institute for Advanced Study and School of Physical Science and Technology, Soochow University, Suzhou 215006, China

[4]Beijing Academy of Quantum Information Sciences, Beijing 100193, China

[5]CAS Center for Excellence in Topological Quantum Computation, University of Chinese Academy of Sciences, Beijing 100190, China

*Corresponding authors: xilin@pku.edu.cn (X. L.); cxc955@psu.edu (C. Z. C.)

**The phase transitions from one plateau to the next plateau or to an insulator in quantum Hall and quantum anomalous Hall (QAH) systems have revealed universal scaling behaviors. A magnetic-field-driven quantum phase transition from a QAH insulator to an axion insulator was recently demonstrated in magnetic topological insulator sandwich samples. Here, we show that the temperature dependence of the derivative of the longitudinal resistance on magnetic field at the transition point follows a characteristic power-law that indicates a universal scaling behavior for the QAH to axion insulator phase transition. Similar to the quantum Hall plateau to plateau transition, the QAH to axion insulator transition can also be understood by the Chalker-Coddington network model. We extract a critical exponent $\kappa \sim 0.38\pm0.02$ in agreement**

**with recent high-precision numerical results on the correlation length exponent of the Chalker-Coddington model at $\nu \sim 2.6$, rather than the generally-accepted value of 2.33.**

The study of quantum phase transitions is a rich topic of research in condensed matter physics. Quantum phase transitions, such as the plateau to plateau transition in the quantum Hall (QH) effect at high magnetic field, can be accessed by varying only one physical parameter (e.g. magnetic field) near absolute zero temperature [1]. The fundamental physics of quantum phase transition can be revealed by investigating its scaling behaviors at the crossing point of magnetoresistance at different temperatures [2, 3, 4, 5]. The universal scaling behavior associated with such a phase transition is usually studied by characterizing a temperature-dependent resistivity/conductivity that is governed by a single exponent $\kappa$ (Refs. [2, 3, 4, 5]). A similar feature of crossing magnetoresistance at different temperatures has also been observed in the QH to Anderson or Hall insulator transition [6, 7, 8, 9, 10].

The quantum anomalous Hall (QAH) state, a zero magnetic field manifestation of the integer QH state, is also a topological phase of quantum matter [11, 12, 13]. Similar to the QH effect [14], the QAH effect harbors dissipationless chiral edge states with quantized Hall resistance and vanishing longitudinal resistance [15]. These edge states are spin-polarized, and their chirality is determined by the internal magnetization of the sample. The QAH effect was envisioned by Duncan Haldane [11] and first realized in magnetically doped topological insulator (TI) thin films [13, 16]. To date, the QAH effect at zero magnetic field has been realized in the Cr- and/or V-doped TI systems [13, 16, 17]. An axion insulator state was recently realized in V-doped TI/TI/Cr-doped TI sandwich structures [18, 19]. When the external magnetic field is swept between the coercive fields of the two magnetic TI films, i.e., the magnetizations of the two surfaces are anti-parallel, the axion insulator state appears. The axion insulator is characterized by zero Hall resistance $R_{xy}$ and a very large longitudinal resistance $R_{xx}$ (Refs. [18, 19]). When the magnetizations of the two magnetic TI layers are driven by an external magnetic field from the antiparallel to parallel alignment, a quantum phase transition from an axion insulator to a QAH insulator has been demonstrated [18, 19].

In this *Article*, we studied the scaling behavior of the quantum phase transition from the QAH insulator to the axion insulator in magnetic TI sandwich samples in the temperature range between 45 and 100 mK. We found the temperature dependence of the derivative of the longitudinal resistance $R_{xx}$ on the magnetic field $B$ evaluated at the critical field $B_c$ follows a characteristic power-law behavior, i.e. $(\frac{dR_{xx}}{dB})_{B=B_c} \propto T^{-\kappa}$. The exponent $\kappa$ is found to be 0.38±0.02, smaller than the previously-believed value $\kappa \sim 0.43$ for the QH plateau to plateau transition [4, 20, 21], but in good agreement with the recent high-precision numerical results on the Chalker-Coddington network model which yields a correlation length exponent $v \sim 2.6$ if the dynamic exponent $p = 2$ (Refs. [22, 23, 24, 25]). The experimental results are supported by theoretical arguments that the QAH to axion insulator phase transition, just like the QH plateau to plateau transition, can also be described using the Chalker-Coddington network model. In addition, by using a quasi-DC measurement circuit, we measured the two-terminal resistance $R_{12}$ in the axion insulator regime and found it to be as high as $5.0 \times 10^4$ $h/e^2$ (~ 1.3 GΩ) (Supplementary Note 1 [26]). This value suggests the highly insulating property of the topology-induced axion insulator state. By analyzing the temperature dependence of $R_{12}$, we probed two different insulating behaviors in two adjacent temperature ranges, validating the reliability of our scaling study for the QAH to axion insulator transition.

## Results

**Sample structures and sample characterizations.** The heterostructure samples used in this work are sandwiches with an undoped TI layer (5QL (Bi, Sb)$_2$Te$_3$ layer) inserted between two magnetic TI layers with a 3QL Cr-doped (Bi, Sb)$_2$Te$_3$ layer in the bottom and a 3QL V-doped (Bi, Sb)$_2$Te$_3$ layer on top (**Fig. 1a**). Specifically, the samples are 3QL (Bi, Sb)$_{1.89}$V$_{0.11}$Te$_3$ /5QL (Bi, Sb)$_2$Te$_3$/ 3QL (Bi, Sb)$_{1.85}$Cr$_{0.15}$Te$_3$ sandwich heterostructures. These samples were grown on 0.5 mm thick heat-treated SrTiO$_3$ (111) substrate in a molecular beam epitaxy (MBE) chamber with a base vacuum of 2×10$^{-10}$ mbar. The Bi/Sb ratio in each layer was optimized to tune the chemical potential near the charge neutral point. The transport studies were performed in two dilution refrigerators (Leiden 6mK/14T and 8mK/9T) with the magnetic field $B$ applied perpendicular to the film plane. We used an excitation current of no more than 0.3 nA with a lowest trusted electron temperature ~ 45 mK (Supplementary Note 2 [26]). Six-terminal Hall bars with a bottom-gate electrode (**Fig. 1b**) were used for electrical transport studies.

**Figure 1c** shows the magnetic field $B$ dependence of the Hall resistance $R_{xy}$ (blue) and the longitudinal resistance $R_{xx}$ (red) of the magnetic TI sandwich sample at $T$ =15 mK and $V_g$ = $V_g^0$ = +21.5 V. Here $V_g^0$ is the charge neutral point, determined by achieving the largest two-terminal resistance in the axion insulator regime. When the magnetization of the bottom Cr-doped TI layer is aligned in parallel with that of the top V-doped TI layer, the sample displays a perfect QAH state: at $B$ = 0 T, $R_{xy}(0)$ = ± $h/e^2$ (within 0.07% of quantization) and $R_{xx}(0)$ < 17 Ω. When the magnetization of the two magnetic TI layers is antiparallel, i.e., $B_{c1} < B < B_{c2}$, where $B_{c1}$ ~ 0.20 T and $B_{c2}$ ~ 0.95 T are the coercive fields of Cr-doped and V-doped TI layers, the sample shows an axion insulator state [19]. Since the axion insulator is a topology-induced insulating state, its $R_{xx}$ value diverges. The traditional four-terminal measurement method is not suitable to measure such a large resistance. The two-terminal method with an AC signal source could deal with larger resistance but the largest resistance is still limited due to the unavoidable parasitic capacitance in the measurement circuit. Neither the traditional four-terminal measurement method nor the two-terminal method with an AC signal source provides reliable resistance values for such an insulating state. Therefore, we show four-terminal data only between 0 to ~10 $h/e^2$, and AC two-terminal data from ~ 10 $h/e^2$ to ~ 1000 $h/e^2$ in **Fig. 1c**. At higher resistance range, we used a quasi-DC measurement scheme. This quasi-DC measurement method can yield reliable two-terminal resistance up to ~ 2 GΩ (~ $7.7 \times 10^4$ $h/e^2$) (Supplementary Note 1 [26]). This method, however, is inconvenient for measurement when sweeping the magnetic field, because it requires very long measurement time and does not provide a "continuous" field-dependent resistance value as the AC lock-in technique.

**QAH insulator to axion insulator quantum phase transition.** When $B > B_{c2} > B_{c1}$, the sample shows the perfect QAH state (**Fig. 1c**), so there must be a magnetic-field-driven quantum phase transition from an axion insulator to a QAH insulator near $B_{c2}$. In order to study the scaling behavior of this quantum phase transition, we measured the magnetic field $B$ dependence of $R_{xx}$ around $B_{c2}$ at different temperatures, as shown in **Fig. 2a**. All $R_{xx}$ - $B$ curves in the temperature range 45 mK ≤ $T$ ≤ 80 mK cross each other at one point, specifically at the critical magnetic field $B = B_c \sim 0.979$ T. $B_c$ separates the $R_{xx}$ - $B$ curves into two regions. For $B < B_c$, $R_{xx}$ increases with lowering temperature, showing an "insulating" behavior. This "insulating" state corresponds to the axion insulator state with a large $R_{xx}$ and zero $R_{xy}$ near-zero temperature [18, 19]. For $B > B_c$, $R_{xx}$ decreases with lowering temperature, exhibiting a "metallic" behavior.

This "metallic" state corresponds to the chiral edge channel of the QAH insulator state with zero $R_{xx}$ and quantized $R_{xy}$ at zero temperature [13, 16]. This "metal" to insulator quantum phase transition is further demonstrated from the temperature dependence of $R_{xx}$ measured at various magnetic fields (**Fig. 2b**). **Figure 2c** shows the value of $R_{xx}$ at $B = B_c$ for different temperatures. The $R_{xx}$ value at $T = 100$ mK shows a clear deviation from the values measured at lower temperatures, which implies that we should confine our investigation of the scaling behavior of the QAH to axion insulator transition in the temperature range of $45 \leq T \leq 80$ mK. Note that one more transition from the QAH insulator to the axion insulator occurs at $B_{c1} \sim 0.20$ T. We experienced instability problem of our magnet at low magnetic field region, so the data acquired near $B_{c1}$ cannot be used for scaling analysis.

**Scaling analysis of the QAH insulator to axion insulator transition.** Next, we analyze the scaling behavior of the QAH to axion insulator phase transition. We follow the scaling analysis of the QH plateau to plateau transition [2, 3] and scale the $B$ dependence of $R_{xx}$ at different temperatures to a single parameter relation: $(\frac{dR_{xx}}{dB})_{B=B_c} \propto T^{-\kappa}$. **Figure 3a** shows the $(\frac{dR_{xx}}{dB})_{B=B_c}$ at different temperatures plotted on log-log scale. The perfect linear fit in the temperature range $45$ mK $\leq T \leq 80$ mK demonstrates the QAH to axion insulator quantum phase transition does show the universal scaling behavior. For $T < 45$ mK, $(\frac{dR_{xx}}{dB})_{B=B_c}$ starts to deviate from the linear dependence, leading us to suspect that the real electron temperature of the sample is higher than the value registered by the thermometer on the mixing chamber in this temperature range (Supplementary Note 2 [26]). We note that $\kappa$ can also be acquired in the $R_{xy}$ analysis over the same temperature range, but it has a lower accuracy due to mutual mixing from huge $R_{xx}$ (Supplementary Note 3 [26]).

From the linear fit in **Fig. 3a**, we extracted a critical exponent $\kappa \sim 0.38\pm0.02$. To further demonstrate the universal scaling behavior of the QAH to axion insulator transition, we used the one-parameter scaling relation $R_{xx} = f[|B-B_c|\cdot T^{-\kappa}]$ to scale the $B$ dependence of $R_{xx}$ at different temperatures as a function of scaled magnetic field $|B-B_c|\cdot T^{-\kappa}$ (Refs. [27, 28]). All curves except the curve at $T$ = 100 mK are well-scaled to a single curve around the critical point. **Figure 3a** shows that including the 100 mK data for the scaling fit does not change $\kappa$, but if more high temperatures data points are included, a larger $\kappa$ is found but not reliable (Supplementary Note 4 [26]). Below, we propose a theoretical picture to explain why the QAH to the axion insulator transition shares the same universality class of QH plateau to plateau transition and also demonstrate that $\kappa \sim 0.38\pm0.02$ is in agreement with the latest high-precision numerical results of the correlation length exponent $\nu \sim 2.6$ (Refs. [22, 23, 24, 25]).

**Theoretical analysis of the QAH insulator to axion insulator transition.** The top and bottom surface states in V-doped TI/TI/Cr-doped TI sandwich heterostructures can be described by the effective Hamiltonian:

$$H_{t,b} = \pm\hbar v_F(s_x k_y - s_y k_x) + [M_{t,b} + \Delta_{t,b}(r)]s_z + V_{t,b}(r). \tag{1}$$

Here $s_{x,y,z}$ are Pauli matrices acting on spin space, $k_{x,y}$ are wave vectors in $x$ and $y$ directions, and $v_F$ is the Fermi velocity. The Dirac mass term $M_{t,b}$ and $\Delta_{t,b}(r)$ represent spatial-averaged and random parts of the exchange field in the $z$-direction due to the magnetization on top and bottom surfaces, respectively, where "$t$" and "$b$" denote the top and bottom surfaces. The random scalar potential $V_{t,b}(r)$ denotes an inhomogeneous onsite potential inside the sandwich sample. In the clean limit, $V_{t,b}(r) = 0$ and $\Delta_{t,b}(r) = 0$, so the Hall conductance of the top and bottom surfaces $\sigma_{xy}^{t,b} = \text{sign}(M_{t,b})e^2/2h$ is determined by the sign of magnetization sign($M_{t,b}$). The QAH phase with $\sigma_{xy}^t = \sigma_{xy}^b$ and the axion insulator phase with $\sigma_{xy}^t = -\sigma_{xy}^b$ require parallel and anti-parallel magnetizations on the top and bottom surfaces, respectively. During the magnetization reversal process, the magnetization of one surface switches direction and a magnetic-field-driven quantum phase transition from the QAH insulator to the axion insulator occurs. Therefore, we consider only one surface in the following and denote it as top surface for concreteness. At the critical point of the QAH insulator to the axion insulator transition, the spatial-averaged magnetization of the top surface $M_t = 0$, whereas $\Delta_t(r) \neq 0$ with an equal population of upward and downward magnetic domains (**Fig. 4a**). Now the top surface state can be described by the random Dirac Hamiltonian with randomness in the mass $\Delta_t(r)$ and the scalar potential $V_t(r)$:

$$H_t = \hbar v_F (s_x k_y - s_y k_x) + \Delta_t(r) s_z + V_t(r) \qquad (2)$$

There exists a chiral edge mode between two regions with opposite masses for the Dirac Hamiltonian [29]. Due to a spatially varying random Dirac mass $\Delta_t(r)$, the random Dirac Hamiltonian $H_t$ exhibits chiral edge modes confined to the zero-mass contours $\Delta_t(r) = 0$ (**Fig. 4a**) [29].

Moreover, it is known that the Chalker-Coddington network model can be mapped to the random Dirac Hamiltonian [29, 30, 31]. **Figure 4b** shows the Chalker-Coddington network model on a square lattice structure. At each node (red dashed square in **Fig. 4b**), a scattering matrix $S$ describes the scattering from incoming (labeled by $Z_{i=1,3}$) to outgoing channels (labeled by $Z_{i=2,4}$) (**Fig. 4c**). Here, the incoming and outgoing channels in the network model correspond to the chiral edge modes at zero-mass domain walls of the random Dirac Hamiltonian (**Fig. 4a**). **Figure 4c** shows the chiral edge channels at the four domains of opposite massed (labeled by $\pm\Delta_t$). All chiral edge channels in the network model can be created similarly and the Chalker-Coddington model is thus equivalent to the random Dirac Hamiltonian [29, 31]. Since the QH plateau to plateau transition (without electron-electron interactions) can also be described by the Chalker-Coddington model [30], the quantum phase transition from the QAH insulator to the axion insulator must share the same universality class of QH plateau to plateau transition. Indeed, the critical exponent $\kappa \sim 0.38\pm0.02$ revealed in our experiment is close to the generally accepted value $\kappa \sim 0.43$ in the QH plateau to plateau transition [4, 20, 21]. The slight discrepancy of $\kappa = p/2\nu$ may result from a decreased dynamic exponent $p$ or an increased correlation exponent $\nu$. However, the theoretical problem of the QH plateau-plateau transition with strong electron-electron interactions remains open and a reliable method for calculating these exponents is yet to be established [32, 33, 34, 35, 36, 37]. Recently, the high-precision numerical calculations show that the correlation length exponent $\nu$ of the Chalker-Coddington model is $\sim 2.6$ (Refs. [22, 23, 24]) instead of $\sim 2.33$. Note that the larger $\nu$ value is also supported by the recent field theory analysis [25]. As a consequence, the critical exponent $\kappa = p/2\nu \sim 0.38$, if we assume $p = 2$ is in good agreement with the $\kappa$ value obtained in our experiment. Here, the exponent $p = 2$, exceeding the non-interacting value $p = 1$, may be induced by the appearance of the Coulomb interaction, similar to the discussion for the QH plateau to plateau transition[32, 33, 34, 35, 36, 37]. In our samples, since the density of states of Dirac fermions is low, the long-ranged Coulomb interaction is unscreened, and the triplet interaction channel is eliminated due to the strong spin-orbital coupling of TI materials [33]. We note that the QAH to axion insulator transition in magnetic TI sandwiches belongs to the QH-type instead of the Berezinskii-Kosterlitz-Thouless-type transition as we predicted in an individual magnetic TI thin film with random domains [38]. In such an individual magnetic TI thin film, the top and bottom surfaces are coupled and play a joint role, so this case is described by the Chalker-Coddington model with two channels [38, 39]. In other words, the magnetization reversal during the QAH plateau to plateau transition occurs on both top and bottom surfaces. This may be the reason why a different $\kappa \sim 0.6$ was found for the QAH to Anderson insulator transition [40] and the QAH plateau to plateau transition [41]. However, the magnetization reversal in magnetic TI sandwich heterostructures during the QAH insulator to the axion insulator transition occurs on *only* one surface.

**Discussion**

A recent study on the quantum phase transition between a Chern insulator to an axion insulator driven by the external magnetic field in even number layers of topological antiferromagnet MnBi$_2$Te$_4$ reports a critical exponent $\kappa \sim 0.47$ (Ref. [42]). In the QH plateau to plateau transition, values of $\kappa$ ranging from 0.15 ~ 0.8 have been reported in the two-dimensional electron/hole gas samples [2, 5, 43, 44, 45, 46, 47, 48, 49, 50]. Several factors can cause larger $\kappa$, for example, the high concentration of ionized impurities or clustered impurities [4]. If the quantum percolation is turning into a classical percolation, $\kappa$ is expected to eventually increase to 0.75 (Refs. [32, 51]). Note that the determination of $\kappa$ is also sensitive to experimental issues, for example, the small temperature regime where all curves crossing each other as shown in **Fig. 2c**. It is not appropriate to include the high-temperature data in our scaling analysis, which can cause a larger and nonreliable $\kappa$ (Supplementary Note 4 [26]). We speculate that the appropriate amount of disorder in our sample might make $\kappa$ as close as to the expected value based on the recent high-precision numerical results of the correlation length exponent $\nu \sim 2.6$ (Refs. [22, 23, 24, 25]).

As noted above, the AC two-terminal measurement is not reliable when $R_{12} > 10^3$ $h/e^2$. In order to measure $R_{12}$ accurately in the most insulating regime, we employed the quasi-DC measurement method. The quasi-DC method uses a low frequency ($f$ = 25.3 mHz) square wave as output, which can eliminate the influence of parasitic capacitance. The thermal voltage can be removed by taking the derivative of the square wave generated current. Therefore, the quasi-DC method can provide more accurate resistance values when $R_{12}$ is between 25.8 MΩ (~ $10^3$ $h/e^2$) and 2 GΩ. By varying the magnetic field $B$, the most insulating state $R_{12,max} \sim 1.3$ GΩ (~ $5.0 \times 10^4$ $h/e^2$) was found at $B$ = 0.40 T at $T$ = 50 mK (Supplementary Note 5 [26]). Since the lowest $R_{xx}$ at QAH state measured at 50 mK was ~ 20 Ω, this suggests an order of ~ $10^8$ ON/OFF resistance ratio.

**Figure 3b** shows the maximum two-terminal resistance $R_{12,max}$ of the axion insulator state measured at different temperatures between 50 mK and 220 mK. The logarithmic $R_{12,max}$ shows a linear dependence on $T^{-1}$ with a slope ~ 333 mK below 95mK, and a steeper slope ~ 478 mK above 95 mK (Supplementary Note 5 [26]). We note that for $T \geq 100$ mK, the scaling behavior starts to disappear (inset of **Fig. 3a**). Therefore, the insulating behavior at the high-temperature regime might involve the contribution from the classical thermal activation, while the low-temperature regime reflects solely the property of an unconventional insulating quantum state (i.e. axion insulator) favoring zero-temperature limit. This further validates our choice to carry out scaling behavior study between 45 mK and 80 mK.

To summarize, we analyzed the critical scaling behavior of the quantum phase transition from a QAH insulator to an axion insulator state in magnetic TI sandwich samples. We found that the derivative of the longitudinal resistance $R_{xx}$ at the critical magnetic field follows a power-law dependence on temperature. Our experimental results combined with theoretical studies on the Chalker-Coddington model demonstrated that the QAH to axion insulator phase transition shares the same universality class with the QH plateau to plateau transition instead of the QAH plateau to plateau transition. Our work opens the door for further explorations of the critical behaviors of quantum phase transitions in topological materials.

**Methods**

**Fabrications of magnetic TI sandwich Hall bar device**

The magnetic TI sandwich heterostructure growth was carried out using a commercial MBE system with a vacuum ~ $2\times10^{-10}$ mbar. The heat-treated insulating SrTiO$_3$ (111) substrates were outgassed at ~ 530 °C for 1 hour before the growth of the TI sandwich heterostructures. High-purity Bi (99.999%), Sb (99.9999%), Cr (99.999%) and Te (99.9999%) were evaporated from Knudsen effusion cells. During the growth of the TI, the substrate was maintained at ~ 240 °C. The flux ratio of Te per (Bi + Sb) was set to be > 10 to prevent Te deficiency in the samples. To avoid possible contamination, a 10 nm thick Te layer is deposited at room temperature on top of the magnetic TI sandwich heterostructures prior to their removal from the MBE chamber. Our prior study has demonstrated than the Te capping layer is much more insulating than the axion insulator in the low-temperature regime and plays a negligible effect in our electrical transport measurements [52]. We next scratched these magnetic TI heterostructure samples into a Hall bar geometry (Supplementary Note 6 [26]) using a computer-controlled probe station. The effective area of the Hall bar devices is ~ 1 mm × 0.5 mm. The electrical ohmic-contacts for dilution measurements were made by pressing indium spheres on the Hall bar. The bottom gate electrode was prepared through an indium foil on the back side of the SrTiO$_3$ substrate.

**Transport measurements.**

Transport measurements were performed using two dilution refrigerators (Leiden 6mK/14T and 8mK/9T) with the magnetic field applied perpendicular to the film plane. The bottom gate voltage was applied using a Keithley 2400. The four-terminal measurements are obtained by applying current from 1 to 2 (**Fig. 1b**), and $R_{xx} = V_{34}/I_{12}$ while $R_{xy} = V_{35}/I_{12}$. The excitation current in the dilution refrigerator measurements is ≤ 0.3 nA to suppress the heating effect. The lowest trusted electron temperature in our dilution transport measurements is ~ 45 mK (Supplementary Note 2 [26]). More details about the dilution measurements can be found in the Supplementary Information.

**Effective model for theoretical analysis.**

In the Chalker-Coddington network model, we consider a square lattice of nodes and there are two incoming and two outgoing channels at each node. Here, the wave function for a particle on each link is represented by the probability amplitudes $Z_{i=1,\ldots,4}$, where a random phase $\varphi_i \in [0, 2\pi]$ is accumulated along with the link. Because of current conservation, $|Z_1|^2 + |Z_3|^2 = |Z_2|^2 + |Z_4|^2$. Therefore, the two incoming channels (labeled by $Z_{i=1,3}$) and two outgoing channels (labeled by $Z_{i=2,4}$) at the node can be related by a 2 × 2 scattering matrix $S$

$$\begin{pmatrix} Z_2 \\ Z_4 \end{pmatrix} = \begin{pmatrix} \cos\alpha & \sin\alpha \\ -\sin\alpha & \cos\alpha \end{pmatrix} \begin{pmatrix} Z_1 \\ Z_3 \end{pmatrix},$$

where $\alpha$ is a real parameter characterizing the tunneling. If $\alpha$ is identical for all nodes, the network model is critical at $\alpha = \alpha_c = \pi/4$. In the experiment, the critical states are described by a random Dirac Hamiltonian with randomness in the mass $\Delta_t(r)$ and the scalar potential $V_t(r)$: $H_t = \hbar v_F (s_x k_y - s_y k_x) + \Delta_t(r) s_z + V_t(r)$. Here the links in the network model are related to the chiral edge modes along contours of zero-mass domain walls, while the scattering parameter $\alpha$ at the nodes can be obtained by solving the eigenvalue equation of $H_t$. In this Hamiltonian, $s_{x,y,z}$ are Pauli matrices acting on spin space, $k_{x,y}$ are wave vectors along $x$ and $y$ directions, and $v_F$ is the Fermi velocity.

**Acknowledgment**

We thank C. X. Liu for helpful discussion. The work in China is supported by the NSFC (11674009), the National Key Research and Development Program of China

(2017YFA0303301), NSFC (11534001 and 110801719), the Beijing Natural Science Foundation (JQ18002), the Strategic Priority Research Program of Chinese Academy of Sciences (Grant XDB28000000), and the Priority Academic Program Development of Jiangsu Higher Education Institutions and NSFC of Jiangsu province (BK20190813). C.-Z. Chang and M.H.W.C. acknowledge the support from the DOE grant (DE-SC0019064). N.S. acknowledges the support from the Penn State 2DCC-MIP under NSF grant DMR-1539916. C.-Z. Chang acknowledges the support from ARO Young Investigator Program Award (W911NF1810198) and the Gordon and Betty Moore Foundation's EPiQS Initiative (Grant GBMF9063 to C.-Z. Chang).

**Author contributions**

C. -Z. Chang and X. L. conceived and designed the experiment. D. X. and L. Z. grew the sandwich heterostructure samples with the help of M. H.W.C., N. S., and C. -Z. Chang. X. W. and J. S. performed the dilution refrigerator measurements with the help of L. X. C.-Z. Chen and X. C. X. provided theoretical support and did all theoretical calculations. X. W., C.-Z. Chen, X. L., and C. -Z. Chang analyzed the data and wrote the manuscript with contributions from all authors.

**Additional information**

Supplementary information is available in the online version of the paper. Reprints and permissions information is available online at www.nature.com/reprints.

Correspondence and requests for materials should be addressed to X. L. or C. -Z. Chang.

**Competing interests**

The authors declare no competing interests.

**Data availability**

The data that support the findings of this study are available from X. L. or C. -Z. Chang upon reasonable request.

**Figures and figure captions**

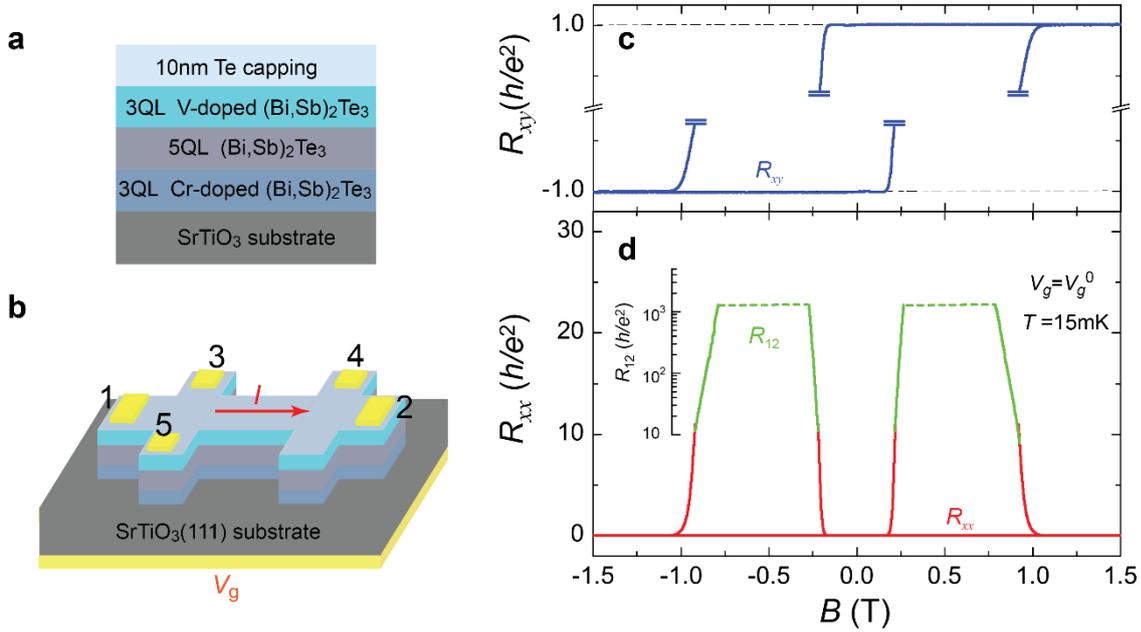

**Figure 1| QAH insulator to axion insulator transition in magnetic TI sandwich heterostructures.** (a) Schematic of the magnetic/nonmagnetic/magnetic TI sandwich heterostructure. (b) Schematic of the Hall bar device used in electrical transport measurements. (c) Magnetic field $B$ dependence of $R_{xy}$ (blue) at $V_g = V_g^0 = +21.5$ V. (d) $B$ dependence of $R_{xx}$ (Red) and $R_{12}$ (green) at $V_g = V_g^0 = +21.5$ V. The four-terminal longitudinal resistance $R_{xx}$ data is shown in red, while the two-terminal resistance $R_{12}$ data is shown in green. When the $R_{12}$ value is comparable with or higher than the impedance of parasitic capacitance of the measurement circuit, the $R_{12}$ values measured with AC two-terminal method are not reliable and are represented by the green dashed lines. $R_{xy}$ values in the highly insulating regime due to the inaccuracy caused by mutual mixing from huge $R_{xx}$ are not shown. Measurements in (c) and (d) were taken when the environmental temperature of the dilution fridge is 15 mK.

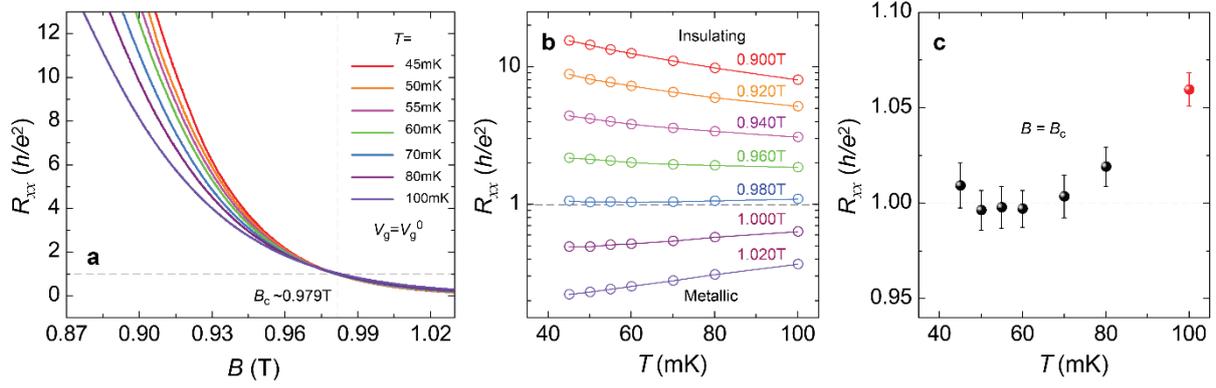

**Figure 2| Temperature evolution of the QAH insulator to axion insulator transition.** (a) Magnetic field dependence of $R_{xx}$ at various temperatures. All curves cross at one critical point at $B_c \sim 0.979$ T. (b) Temperature dependence of $R_{xx}$ at different magnetic fields. The sample shows insulating behavior for $B < B_c$ and "metallic" behavior for $B > B_c$. At $B = 0.980$ T, $R_{xx}$ is nearly independent of temperature. (c) $R_{xx}$ value at the crossing point as a function of temperature.

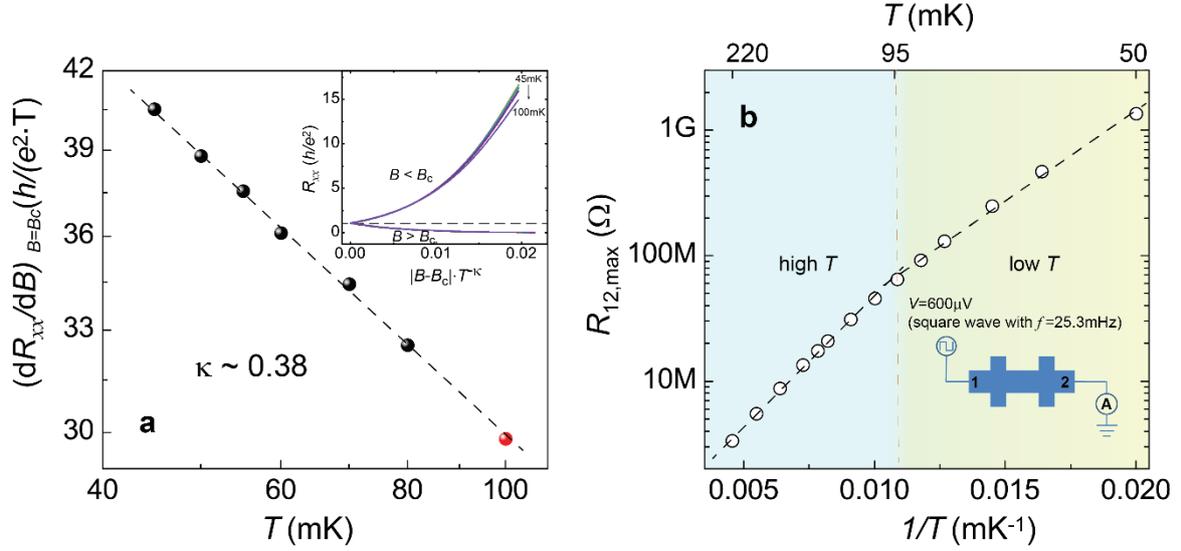

**Figure 3| Scaling behavior of the QAH insulator to axion insulator transition.** (a) Temperature dependence of $(\frac{dR_{xx}}{dB})_{B=B_c}$ on log-log scale. The black dashed line is the linear fit. The inset shows $R_{xx}$ as a function of the scaled magnetic field $|B-B_c|\cdot T^{-\kappa}$ at different temperatures. (b) The two-terminal DC resistance $R_{12}$ as a function of $1/T$ (bottom axis) or $T$ (top axis) when the sample shows the most insulating state (i.e. $B = 0.40$ T). The vertical axis is on a log scale. The inset shows the measurement circuit we used to accurately measure the huge $R_{12}$ of the axion insulator state.

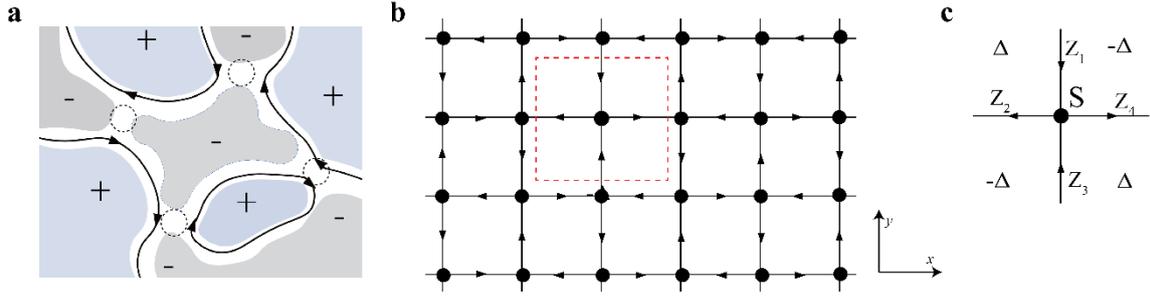

**Figure 4| Chiral edge modes along magnetic domain walls at the transition regime from the QAH insulator to the axion insulator and the Chalker-Coddington network model.** (a) The chiral zero-energy edge states along contours of zero-mass domain walls of Dirac fermions. "+" and "-" denote domains of positive and negative masses of Dirac fermions, respectively. (b) The Chalker-Coddington network model on a square lattice. (c) At each node (red dashed square in b), a scattering matrix $S$ describes the scattering from incoming (labeled by $Z_{i=1,3}$) to outgoing modes (labeled by $Z_{i=2,4}$). The chiral edge modes can be created by Dirac Hamiltonian with four domains of opposite Dirac masses labeled by $\pm\Delta$.

**References**

1. Sondhi S. L., Girvin S. M., Carini J. P., Shahar D. Continuous quantum phase transitions. *Rev. Mod. Phys.* **69**, 315-333 (1997).

2. Wei H. P., Tsui D. C., Paalanen M. A., Pruisken A. M. M. Experiments on Delocalization and Universality in the Integral Quantum Hall-Effect. *Phys. Rev. Lett.* **61**, 1294-1296 (1988).

3. Pruisken A. M. M. Universal Singularities in the Integral Quantum Hall-Effect. *Phys. Rev. Lett.* **61**, 1297-1300 (1988).

4. Li W. L., Csathy G. A., Tsui D. C., Pfeiffer L. N., West K. W. Scaling and universality of integer quantum Hall plateau-to-plateau transitions. *Phys. Rev. Lett.* **94**, 206807 (2005).

5. Li W. L., Vicente C. L., Xia J. S., Pan W., Tsui D. C., Pfeiffer L. N., West K. W. Scaling in Plateau-to-Plateau Transition: A Direct Connection of Quantum Hall Systems with the Anderson Localization Model. *Phys. Rev. Lett.* **102**, 216801 (2009).

6. Jiang H. W., Johnson C. E., Wang K. L., Hannahs S. T. Observation of Magnetic-Field-Induced Delocalization - Transition from Anderson Insulator to Quantum Hall Conductor. *Phys. Rev. Lett.* **71**, 1439-1442 (1993).

7. Shahar D., Tsui D. C., Shayegan M., Bhatt R. N., Cunningham J. E. Universal Conductivity at the Quantum Hall Liquid to Insulator Transition. *Phys. Rev. Lett.* **74**, 4511-4514 (1995).

8. Song S. H., Shahar D., Tsui D. C., Xie Y. H., Monroe D. New universality at the magnetic field driven insulator to integer quantum Hall effect transitions. *Phys. Rev. Lett.* **78**, 2200-2203 (1997).

9. van Schaijk R. T. F., de Visser A., Olsthoorn S. M., Wei H. P., Pruisken A. M. M. Probing the plateau-insulator quantum phase transition in the quantum Hall regime. *Phys. Rev. Lett.* **84**, 1567-1570 (2000).

10. Smorchkova I. P., Samarth N., Kikkawa J. M., Awschalom D. D. Giant magnetoresistance and quantum phase transitions in strongly localized magnetic two-dimensional electron gases. *Phys. Rev. B* **58**, R4238-R4241 (1998).

11. Haldane F. D. M. Model for a Quantum Hall-Effect without Landau Levels: Condensed-Matter Realization of the "Parity Anomaly". *Phys. Rev. Lett.* **61**, 2015-2018 (1988).

12. Yu R., Zhang W., Zhang H. J., Zhang S. C., Dai X., Fang Z. Quantized Anomalous Hall

Effect in Magnetic Topological Insulators. *Science* **329**, 61-64 (2010).

13. Chang C. Z., Zhang J. S., Feng X., Shen J., Zhang Z. C., Guo M. H., Li K., Ou Y. B., Wei P., Wang L. L., Ji Z. Q., Feng Y., Ji S. H., Chen X., Jia J. F., Dai X., Fang Z., Zhang S. C., He K., Wang Y. Y., Lu L., Ma X. C., Xue Q. K. Experimental Observation of the Quantum Anomalous Hall Effect in a Magnetic Topological Insulator. *Science* **340**, 167-170 (2013).

14. von Klitzing K., Dorda G., Pepper M. New Method for High-Accuracy Determination of the Fine-Structure Constant Based on Quantized Hall Resistance. *Phys. Rev. Lett.* **45**, 494-497 (1980).

15. Chang C. Z., Zhao W. W., Kim D. Y., Wei P., Jain J. K., Liu C. X., Chan M. H. W., Moodera J. S. Zero-Field Dissipationless Chiral Edge Transport and the Nature of Dissipation in the Quantum Anomalous Hall State. *Phys. Rev. Lett.* **115**, 057206 (2015).

16. Chang C. Z., Zhao W. W., Kim D. Y., Zhang H. J., Assaf B. A., Heiman D., Zhang S. C., Liu C. X., Chan M. H. W., Moodera J. S. High-Precision Realization of Robust Quantum Anomalous Hall State in a Hard Ferromagnetic Topological Insulator. *Nat. Mater.* **14**, 473-477 (2015).

17. Ou Y., Liu C., Jiang G., Feng Y., Zhao D., Wu W., Wang X. X., Li W., Song C., Wang L. L., Wang W., Wu W., Wang Y., He K., Ma X. C., Xue Q. K. Enhancing the Quantum Anomalous Hall Effect by Magnetic Codoping in a Topological Insulator. *Adv. Mater.* **30**, 1703062 (2017).

18. Mogi M., Kawamura M., Tsukazaki A., Yoshimi R., Takahashi K. S., Kawasaki M., Tokura Y. Tailoring Tricolor Structure of Magnetic Topological Insulator for Robust Axion Insulator. *Sci. Adv.* **3**, eaao1669 (2017).

19. Xiao D., Jiang J., Shin J. H., Wang W. B., Wang F., Zhao Y. F., Liu C. X., Wu W. D., Chan M. H. W., Samarth N., Chang C. Z. Realization of the Axion Insulator State in Quantum Anomalous Hall Sandwich Heterostructures. *Phys. Rev. Lett.* **120**, 056801 (2018).

20. Huckestein B. One-Parameter Scaling in the Lowest Landau Band. *Physica A* **167**, 175-187 (1990).

21. Huckestein B. Scaling Theory of the Integer Quantum Hall-Effect. *Rev. Mod. Phys.* **67**, 357-396 (1995).

22. Slevin K., Ohtsuki T. Critical exponent for the quantum Hall transition. *Phys. Rev. B* **80**, 041304 (2009).

23. Amado M., Malyshev A. V., Sedrakyan A., Dominguez-Adame F. Numerical Study of the Localization Length Critical Index in a Network Model of Plateau-Plateau Transitions in the Quantum Hall Effect. *Phys. Rev. Lett.* **107**, 066402 (2011).

24. Obuse H., Gruzberg I. A., Evers F. Finite-Size Effects and Irrelevant Corrections to Scaling Near the Integer Quantum Hall Transition. *Phys. Rev. Lett.* **109**, 206804 (2012).

25. Zirnbauer M. R. The integer quantum Hall plateau transition is a current algebra after all. *Nucl. Phys. B* **941**, 458-506 (2019).

26. See Supplementary Information for further details regarding the quasi-DC measurement method, estimation of the electron temperature, scaling analysis based on Hall resistance, scaling analysis on two more samples, and other supporting data.

27. Wang T., Clark K. P., Spencer G. F., Mack A. M., Kirk W. P. Magnetic-Field-Induced Metal-Insulator-Transition in 2 Dimensions. *Phys. Rev. Lett.* **72**, 709-712 (1994).

28. Shahar D., Tsui D. C., Shayegan M., Shimshoni E., Sondhi S. L. A different view of the quantum Hall plateau-to-plateau transitions. *Phys. Rev. Lett.* **79**, 479-482 (1997).

29. Ludwig A. W. W., Fisher M. P. A., Shankar R., Grinstein G. Integer Quantum Hall Transition - an Alternative Approach and Exact Results. *Phys. Rev. B* **50**, 7526-7552 (1994).

30. Chalker J. T., Coddington P. D. Percolation, Quantum Tunnelling and the Integer Hall-Effect. *J. Phys. C. Solid State* **21**, 2665-2679 (1988).

31. Ho C. A., Chalker J. T. Models for the integer quantum hall effect: The network model, the Dirac equation, and a tight-binding Hamiltonian. *Phys. Rev. B* **54**, 8708-8713 (1996).

32. Lee D. H., Wang Z. Q., Kivelson S. Quantum Percolation and Plateau Transitions in the Quantum Hall-Effect. *Phys. Rev. Lett.* **70**, 4130-4133 (1993).

33. Belitz D., Kirkpatrick T. R. The Anderson-Mott Transition. *Rev. Mod. Phys.* **66**, 261-390 (1994).

34. Yang S. R. E., Macdonald A. H. Coulomb Gaps in a Strong Magnetic-Field. *Phys. Rev. Lett.* **70**, 4110-4113 (1993).

35. Lee D. H., Wang Z. Q. Effects of electron-electron interactions on the integer quantum Hall transitions. *Phys. Rev. Lett.* **76**, 4014-4017 (1996).

36. Wang Z. Q., Fisher M. P. A., Girvin S. M., Chalker J. T. Short-range interactions and scaling near integer quantum Hall transitions. *Phys. Rev. B* **61**, 8326-8333 (2000).

37. Burmistrov I. S., Bera S., Evers F., Gornyi I. V., Mirlin A. D. Wave function multifractality and dephasing at metal-insulator and quantum Hall transitions. *Annals of Physics* **326**, 1457-1478 (2011).

38. Chen C. Z., Liu H. W., Xie X. C. Effects of Random Domains on the Zero Hall Plateau in the Quantum Anomalous Hall Effect. *Phys. Rev. Lett.* **122**, 026601 (2019).

39. Xiong G., Wang S. D., Niu Q., Tian D. C., Wang X. R. Metallic phase in quantum Hall systems due to inter-Landau-band mixing. *Phys. Rev. Lett.* **87**, 216802 (2001).

40. Chang C. Z., Zhao W. W., Li J., Jain J. K., Liu C. X., Moodera J. S., Chan M. H. W. Observation of the Quantum Anomalous Hall Insulator to Anderson Insulator Quantum Phase Transition and its Scaling Behavior. *Phys. Rev. Lett.* **117**, 126802 (2016).

41. Kawamura M., Mogi M., Yoshimi R., Tsukazaki A., Kozuka Y., Takahashi K. S., Kawasaki M., Tokura Y. Topological quantum phase transition in magnetic topological insulator upon magnetization rotation. *Phys. Rev. B* **98**, 140404 (2018).

42. Liu C., Wang Y. C., Li H., Wu Y., Li Y. X., Li J. H., He K., Xu Y., Zhang J. S., Wang Y. Y. Robust axion insulator and Chern insulator phases in a two-dimensional antiferromagnetic topological insulator. *Nat. Mater.* **19**, 522-527 (2020).

43. Wakabayashi J., Yamane M., Kawaji S. Experiments on the Critical Exponent of Localization in Landau Subbands with the Landau Quantum Numbers 0 and 1 in Si-Mos Inversion-Layers. *J. Phys. Soc. Jpn.* **58**, 1903-1905 (1989).

44. Koch S., Haug R. J., Vonklitzing K., Ploog K. Experiments on Scaling in Alxga1-Xas/Gaas Heterostructures under Quantum Hall Conditions. *Phys. Rev. B* **43**, 6828-6831 (1991).

45. Koch S., Haug R. J., Vonklitzing K., Ploog K. Size-Dependent Analysis of the Metal-Insulator-Transition in the Integral Quantum Hall-Effect. *Phys. Rev. Lett.* **67**, 883-886 (1991).

46. Wei H. P., Lin S. Y., Tsui D. C., Pruisken A. M. M. Effect of Long-Range Potential Fluctuations on Scaling in the Integer Quantum Hall-Effect. *Phys. Rev. B* **45**, 3926-3928 (1992).

47. Koch S., Haug R. J., Vonklitzing K., Ploog K. Experimental Studies of the Localization Transition in the Quantum Hall Regime. *Phys. Rev. B* **46**, 1596-1602 (1992).

48. Liu C. X., Zhang H., Yan B. H., Qi X. L., Frauenheim T., Dai X., Fang Z., Zhang S. C. Oscillatory Crossover from Two-Dimensional to Three-Dimensional Topological

Insulators. *Phys. Rev. B* **81**, 041307 (2010).

49. Wang X. B., Liu H. W., Zhu J. B., Shan P. J., Wang P. J., Fu H. L., Du L. J., Pfeiffer L. N., West K. W., Xie X. C., Du R. R., Lin X. Scaling properties of the plateau transitions in the two-dimensional hole gas system. *Phys. Rev. B* **93**, 075307 (2016).

50. Shan P. J., Fu H. L., Wang P. J., Yang J. X., Pfeiffer L. N., West K. W., Lin X. Coherence length saturation at the low temperature limit in two-dimensional hole gas. *Physica E* **99**, 118-122 (2018).

51. Trugman S. A. Localization, Percolation, and the Quantum Hall-Effect. *Phys. Rev. B* **27**, 7539-7546 (1983).

52. Zhang J. S., Chang C. Z., Zhang Z. C., Wen J., Feng X., Li K., Liu M. H., He K., Wang L. L., Chen X., Xue Q. K., Ma X. C., Wang Y. Y. Band Structure Engineering in $(Bi_{1-x}Sb_x)_2Te_3$ Ternary Topological Insulators. *Nat. Commun.* **2**, 574 (2011).